\documentclass[a4paper,fleqn]{cas-sc}

\usepackage[numbers]{natbib}

\def\tsc#1{\csdef{#1}{\textsc{\lowercase{#1}}\xspace}}
\tsc{WGM}
\tsc{QE}
\tsc{EP}
\tsc{PMS}
\tsc{BEC}
\tsc{DE}

\begin{document}
\let\WriteBookmarks\relax
\def\floatpagepagefraction{1}
\def\textpagefraction{.001}

\shorttitle{}

\shortauthors{X. Jian et~al.}

\title [mode = title]{Modal Decomposition and Identification for a Population of Structures Using Physics-Informed Graph Neural Networks and Transformers}



%
\author[1,2]{Xudong Jian}[orcid=0000-0001-9973-9662]



\ead{xudong.jian@sec.ethz.ch}


\credit{Conceptualization, Methodology, Software, Data curation, Formal analysis, Data visualization, Writing - original draft, Writing - Review \& Editing}

\affiliation[1]{organization={Future Resilient Systems, Singapore-ETH Centre},
    city={Singapore},
    postcode={138602}, 
    country={Singapore}}

\author[1,2]{Kiran Bacsa}[orcid=0000-0002-0834-185X]
\credit{Methodology, Software, Writing - Review \& Editing}
\affiliation[2]{organization={Department of Civil, Environmental and Geomatic Engineering},
    addressline={ETH Zurich}, 
    city={Zurich},
    postcode={8049}, 
    country={Switzerland}}

\author[2]{Gregory Duthé}[orcid=0000-0002-0895-6766]
\credit{Methodology, Software, Writing - Review \& Editing}

\author[1,2]{Eleni Chatzi}[orcid=0000-0002-6870-240X]
\cormark[1]
\ead{chatzi@ibk.baug.ethz.ch}
\credit{Supervision, Funding Acquisition, Methodology, Project administration, Writing - Review \& Editing}

\begin{abstract}
Modal identification is crucial for structural health monitoring and structural control, providing critical insights into structural dynamics and performance. This study presents a novel deep learning framework that integrates graph neural networks (GNNs), transformers, and a physics-informed loss function to achieve modal decomposition and identification across a population of structures. The transformer module decomposes multi-degrees-of-freedom (MDOF) structural dynamic measurements into single-degree-of-freedom (SDOF) modal responses, facilitating the identification of natural frequencies and damping ratios. Concurrently, the GNN captures the structural configurations and identifies mode shapes corresponding to the decomposed SDOF modal responses. The proposed model is trained in a purely physics-informed and unsupervised manner, leveraging modal decomposition theory and the independence of structural modes to guide learning without the need for labeled data. Validation through numerical simulations and laboratory experiments demonstrates its effectiveness in accurately decomposing dynamic responses and identifying modal properties from sparse structural dynamic measurements, regardless of variations in external loads or structural configurations. Comparative analyses against established modal identification techniques and model variations further underscore its superior performance, positioning it as a favorable approach for population-based structural health monitoring.
\end{abstract}


\begin{highlights}
\item \textbf{Population-based modal identification framework}. The proposed approach is tailored for population-based structural health monitoring (PBSHM), enabled by Graph Neural Networks (GNNs) that effectively model structural topologies with varying node configurations. This facilitates scalable modal identification across multiple structures under diverse loading conditions.
\item \textbf{Label-free modal decomposition guided by physics}. By integrating Set Transformers and a physics-informed loss function grounded in modal decomposition theory, the model performs accurate decomposition of dynamic responses into single-degree-of-freedom (SDOF) modes without requiring labeled data, significantly enhancing practical applicability in real-world operational modal analysis (OMA).
\item \textbf{Full-field identification from sparse measurements}. A feature propagation algorithm leverages structural topology to reconstruct full-field mode shapes and dynamic responses from sparse sensor data, addressing a key limitation in conventional SHM where measurements are often incomplete.
\item \textbf{Superior performance under realistic conditions}. Validated through both numerical simulations and laboratory experiments, the proposed model outperforms traditional modal identification methods (e.g., EFDD, SSI) and standard deep learning models (e.g., MLPs, LSTMs) in identifying natural frequencies, damping ratios, and high-order mode shapes, particularly under non-stationary and data-sparse conditions.
\end{highlights}

\begin{keywords}
population-based SHM \sep modal decomposition \sep modal identification \sep physics-informed neural networks \sep graph neural networks \sep transformers \sep
\end{keywords}

\maketitle

\section{Introduction}
\subsection{Motivation}
Structural health monitoring (SHM) and vibration control are essential for ensuring the safety, functionality, and resilience of civil structures such as buildings, bridges, and wind turbines. As civil structures are increasingly subjected to dynamic loads such as traffic-induced vibrations, earthquakes, wind, and waves, the need for SHM and structural vibration control has become more critical. SHM enables real-time assessment of structural integrity, allowing for early detection of damage and deterioration, which is crucial for preventing catastrophic failures \cite{farrar2007introduction}. In parallel, structural vibration control techniques, including passive, active, and hybrid approaches, have been developed to mitigate excessive vibrations, enhancing the resilience and performance of structures under unwanted dynamic loading conditions \cite{spencer2003state}.

A fundamental aspect of both SHM and structural vibration control is the ability to accurately capture and interpret the dynamic behavior of structures \cite{Ibrahim_2017,goyal2016vibration}. Modal decomposition and identification play a pivotal role in this process, as they provide insight into the modal properties of a structure \cite{lai2022neural}, including natural frequencies, mode shapes, damping characteristics, and modal response. For SHM, changes in these modal properties could serve as indicators of structural damage, making modal identification a widely-accepted tool for damage detection and health assessment \cite{osti_249299,nagarajaiah2009output,an2019recent}. Moreover, structural vibration control strategies generally rely on precise modal information to optimize control algorithms, ensuring effective vibration mitigation and stability enhancement \cite{singh2003efficient,braghin2013new,sun2017cables}. Since modern civil structures become more complex and exposed to increasing risks, the development of more advanced modal decomposition and identification methods remains a critical area of research.

\subsection{Related Work}
There have been extensive studies in modal identification for structures, which can be generally classified into Experimental Modal Analysis (EMA) or Operational Modal Analysis (OMA) techniques. Early modal identification techniques fall under EMA, which was developed over half a century ago to identify structural modal properties by applying controlled excitations (such as impact hammers or shakers) and measuring the response \cite{avitabile2001experimental}. While effective in laboratory settings, EMA is often impractical for real-world civil structures due to inevitable environmental noise and the high costs associated with generating artificial excitation for large-scale systems. To overcome these limitations, OMA techniques were developed since 1980s to identify modal properties under actual operational conditions \cite{zahid2020review}. By capturing structural vibration responses induced by ambient sources such as wind and traffic, OMA avoids the need for expensive and intrusive testing, making it particularly valuable for large civil infrastructure. In this sense, OMA is also called output-only modal identification, as it only uses output responses without measuring input forces. Moreover, OMA techniques are generally categorized into frequency-domain and time-domain approaches, with Frequency Domain Decomposition (FDD) \cite{brincker2000modal} and Stochastic Subspace Identification (SSI) \cite{peeters1999reference} being the most widely used techniques in their respective domains. However, classic OMA techniques still face challenges in mode separation, automation, and robustness against noise and environmental variability. To address these challenges, many advanced OMA techniques have been developed over the past two decades, with a notable class of methods integrating signal decomposition techniques, such as independent component analysis (ICA) \cite{yang2013time} and variational model decomposition (VMD) \cite{bagheri2018structural}, for modal identification. The core idea behind these approaches is to decompose the dynamic measurements of multi-degree-of-freedom (MDOF) structures into the dynamic responses of single-degree-of-freedom (SDOF) vibration modes while isolating measurement noise \cite{antoni2005blind}. This not only simplifies modal identification, as identifying modal properties of SDOF systems is more straightforward, but also provides the dynamic response of structural vibration modes as a valuable byproduct, enabling applications in vibration control and structural design optimization. Nonetheless, these techniques still demonstrate limitations such as inadequate mode separation (mode mixing), end effects, lack of physical interpretability, and dependence on decomposition parameters \cite{sadhu2017review}.

Modal identification, like other inverse problems, is essentially an optimization problem that seeks to minimize the discrepancy between identified modal properties and their true values. As a powerful tool to solve complex optimization problems, a few studies have recently introduced machine learning and deep learning approaches to enhance modal decomposition and identification. Based on the model training approach, these studies can be categorized into two types: 1) \textit{Unsupervised learning-based approaches}: Liu et al. \cite{liu2021machine} were among the first to apply machine learning to output-only modal decomposition and identification. By designing a loss function that leverages the independent characteristics of structural modes, they trained a neural network (NN) without labeled data, transforming the training process into a modal decomposition and identification procedure. Building on the work of Liu et al.\cite{liu2021machine}, Bao et al. \cite{bao2024mechanics} incorporated additional physical properties of structural vibration modes and proposed a mechanics-informed NN. This approach enables adaptive mode order determination, spurious mode removal, and automatic modal parameter identification. Both studies were validated using numerical simulations and real-world bridge SHM datasets. The results demonstrated that the proposed approaches achieved comparable accuracy and greater robustness than conventional modal identification techniques, including FDD, SSI, and ICA. However, neither study examined the generalization capability of their trained models for unseen time series or structures. This means the modal identification can only be done through model training, leading to excessive identification time. 2) \textit{Supervised learning-based approaches:} Shu et al. \cite{shu2023multi} proposed a multi-task deep NN for modal decomposition and identification, where sparse component analysis (SCA) was employed to decompose measured structural dynamic responses for training set preparation. González et al. \cite{hernandez2024ai} further advanced this approach by developing a multitask learning deep NN capable of identifying both the real and imaginary components of structural modes, utilizing second-order blind identification (SOBI) to generate training data. Validation on unseen numerical and real-world monitoring datasets demonstrated the automation, generalization ability, and efficiency of these deep NNs in modal decomposition and identification. However, both studies essentially trained models to replicate the behavior of SCA and SOBI. As a result, while the computational efficiency was significantly improved, the accuracy of the trained NNs could not exceed that of SCA and SOBI. Additionally, since the training datasets were derived from a single structure, the generalization ability of these models to new structures remains uncertain.

Recently, a new SHM paradigm, known as population-based SHM (PBSHM), has been develop to enhance the generalization ability of data-driven SHM methods \cite{worden2020abrief}. Unlike traditional SHM methods that focus on individual structures, PBSHM considers a population of structures, in which each member shares certain topological characteristics but differs in geometric or material properties, to enable learning and knowledge transfer among population members \cite{gosliga2021foundations,tsialiamanis2021foundations,tsialiamanis2023towards}. Inspired by PBSHM, our previous study \cite{jian2024using} proposed an OMA approach that leverages graph neural networks (GNN) to identify modal properties of a population of structures. This approach proposed a GNN-based deep learning model and trains it using acceleration power spectral density (PSD) data and the corresponding modal properties of a structural population. Once trained, the model can accurately and efficiently identify the modal properties of unseen structures. However, the practicality of this approach is limited by its supervised learning nature, as it requires prior knowledge of the modal properties of structures, which is typically unavailable in practice, before it can be trained for modal property identification. Additionally, this approach can only identify absolute mode shapes due to its reliance on PSD data, which does not include phase information. Despite these limitations, this study still represents a significant breakthrough, enabling deep learning-based OMA to be generalized across different structures.

\subsection{Our Contributions}
Building on the aforementioned background, this study proposes a novel deep learning-based approach for modal decomposition and identification in a population of structures. The key contributions of this research are as follows:

1) A deep learning model integrating GNNs and transformers is developed. This model decomposes the dynamic measurements of a population of MDOF structures into the time-domain dynamic responses of SDOF vibration modes and identifies the corresponding mode shapes. The extracted SDOF modal responses are further analyzed to identify natural frequencies and damping ratios.

2) A feature propagation algorithm is employed to utilize the structural topology information, which is often overlooked by conventional modal identification methods. This enables the reconstruction of full-field acceleration measurements from sparse data collected at a limited number of structural nodes, alleviating the common challenge in SHM where only partial dynamic response measurements are available.

3) A physics-informed loss function is introduced, leveraging modal decomposition theory and the independent characteristics of structural vibration modes. This eliminates the need for labeled data, significantly enhancing the practical applicability of the proposed approach in real-world OMA.

\section{Methodology}
\subsection{Problem Statement}
\label{sec:problem}
This study aims to achieve modal decomposition and identification of linear structures under realistic conditions, where the structural loads are unknown, and only a limited number of structural nodes are monitored. To formally define the problem, we denote the available dynamic measurements of structures as $\mathbf{X} \in \mathbb{R}^{M \times T}$, where $M$ denotes the amount of monitored nodes, and $T$ denotes the amount of available time samples per signal. For existing modal identification methods, the modal identification process can be expressed as:

\begin{equation}
\left[ \hat{\mathbf{F}}, \hat{\mathbf{Z}}, \hat{\mathbf{\Phi}} \right] = \mathcal{M}(\mathbf{X})
\end{equation}
where $\mathcal{M}(\cdot)$ denotes the modal identification method. $\hat{\mathbf{F}} \in \mathbb{R}_{>0}^{1 \times P}$, $\hat{\mathbf{Z}} \in \mathbb{R}_{>0}^{1 \times P}$, and $\hat{\mathbf{\Phi}} \in \mathbb{R}^{M \times P}$ denote identified natural frequencies (in Hz), damping ratios, and mode shapes of $P$ identified structural vibration modes, respectively.

In practice, the number of structural nodes significantly exceeds the number of measurement points, meaning that $M$ is typically much smaller than the total number of $N$ structural nodes. However, it is always desirable to identify mode shapes on all structural nodes. To do so, this study adopts the feature propagation (FP) algorithm , which estimates the unknown measurements based on the known data, thereby obtaining structural dynamic measurements for all structural nodes. The feature propagation process is formulated as:

\begin{equation}
\tilde{\mathbf{X}} = \mathcal{P}(\mathbf{X})
\end{equation}
where $\mathcal{P}(\cdot)$ represents the FP algorithm, and $\tilde{\mathbf{X}} \in \mathbb{R}^{N \times T}$ denotes the reconstructed dynamic measurements on all $N$ structural nodes.

We then propose a deep learning model to learn a function $f_{\theta}\left ( \cdot  \right )$, where $\theta$ are learnable parameters, to decompose MDOF dynamic measurements $\tilde{\mathbf{X}}$ into SDOF dynamic response of structural vibration modes and identify corresponding mode shapes. This modal decomposition process can be expressed as:

\begin{equation}
\left[ \hat{\mathbf{Q}}, \hat{\mathbf{\Phi}} \right] = f_{\theta}(\tilde{\mathbf{X}})
\end{equation}
where $\hat{\mathbf{Q}}\in\mathbb{R}^{P\times T}$ denotes the time-domain dynamic response of the $P$ decomposed structural modes, and $\hat{\mathbf{\Phi}}\in\mathbb{R}^{N\times P}$ represents the corresponding identified mode shapes on all $N$ structural nodes. Based on $\hat{\mathbf{Q}}$, we further adopt well-established methods, including power spectral density (PSD) analysis and random decrement technique (RDT) \cite{ibrahim1977random}, to identify natural frequencies $\hat{\mathbf{F}}$ and damping ratios $\hat{\mathbf{Z}}$ of  decomposed modes, respectively.

\subsection{Feature Propagation}
\label{sec:FP}
In this study, we adopt the feature propagation (FP) algorithm \cite{rossi2022unreasonable} to reconstruct dynamic measurements for all structural nodes using the available dynamic measurements from a small subset of nodes and the structure's topology. Given a structure with $N$ nodes in total, its topology can be represented by a symmetric $N \times N$ adjacency matrix $\mathbf{A}$, which consists of binary values $A_{ij}$ that indicate the connection between structural nodes $i$ and $j$. Specifically, $A_{ij}=1$ means there is an edge (element) between two nodes, whereas $A_{ij}=0$ indicates no direct connection. 
The FP algorithm calculates the normalized adjacency matrix, which is defined as $\tilde{\mathbf{A}}=\mathbf{D}^{-\frac{1}{2}}\mathbf{A}\mathbf{D}^{\frac{1}{2}}$, where $\mathbf{D}=\text{diag}\left( \sum_{j}^{N}A_{1j},\ldots,\sum_{j}^{N}A_{Nj} \right)$ is the diagonal degree matrix. Based on these preliminaries and permutation invariance of graph-structured data, the FP algorithm further defines:

\begin{equation}
\begin{matrix} 
\mathbf{X}(t)=\left[\begin{matrix}\mathbf{X}_k(t)\\\mathbf{X}_u(t)\\\end{matrix}\right] & \mathbf{A}=\left[\begin{matrix}\mathbf{A}_{kk}&\mathbf{A}_{ku}\\\mathbf{A}_{uk}&\mathbf{A}_{uu}\\\end{matrix}\right] 
\end{matrix}
\end{equation}
where the subscript $u$ and $k$ denote structural nodes with unknown and known measurements, respectively.

Let $\mathbf{X}^{(0)}$ denote the original incomplete dynamic measurements on all structural nodes, in which unknown measurements $\mathbf{X}_u$ are initialized to 0. The FP algorithm propagates known measurements to structural nodes lacking measurements in an iterative way:

\begin{equation}
\label{eq:FP}
\mathbf{X}^{\left(n\right)}=\left[\begin{matrix}\mathbf{I}&\mathbf{0}\\\tilde{\mathbf{A}}_{uk}&\tilde{\mathbf{A}}_{uu}\\\end{matrix}\right]\mathbf{X}^{\left(n-1\right)}
\end{equation}
where $\mathbf{I}$ is an identity matrix.


According to Rossi et al. \cite{rossi2022unreasonable}, 40 iterations are enough to provide convergence when using Equation \eqref{eq:FP} to reconstruct dynamic measurements on all structural nodes. Therefore, in this study we use $\tilde{\mathbf{X}}=\mathbf{X}^{\left(40\right)}$ for modal decomposition and identification.

\subsection{Model Architecture}
\label{sec:architecture}
\begin{figure}[!htp] 
 \centering
 \includegraphics[scale=0.7]{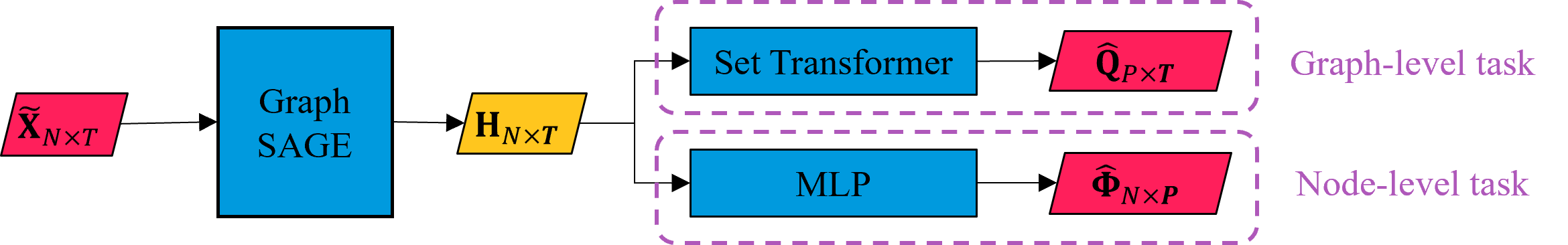}
 \caption{Architecture of the proposed model, in which the input and output are highlighted in red, hidden features in yellow, and deep learning blocks in blue.}
    \label{fig:architecture}
\end{figure}

After using the FP algorithm to obtain estimated complete dynamic measurements $\tilde{\mathbf{X}}$, we design a deep learning model to learn the mapping between $\tilde{\mathbf{X}}$ and $\left[ \hat{\mathbf{Q}}, \hat{\mathbf{\Phi}} \right]$. The architecture of the proposed model is shown in Figure \ref{fig:architecture}.

\begin{figure}[!htp] 
 \centering
 \includegraphics[scale=0.8]{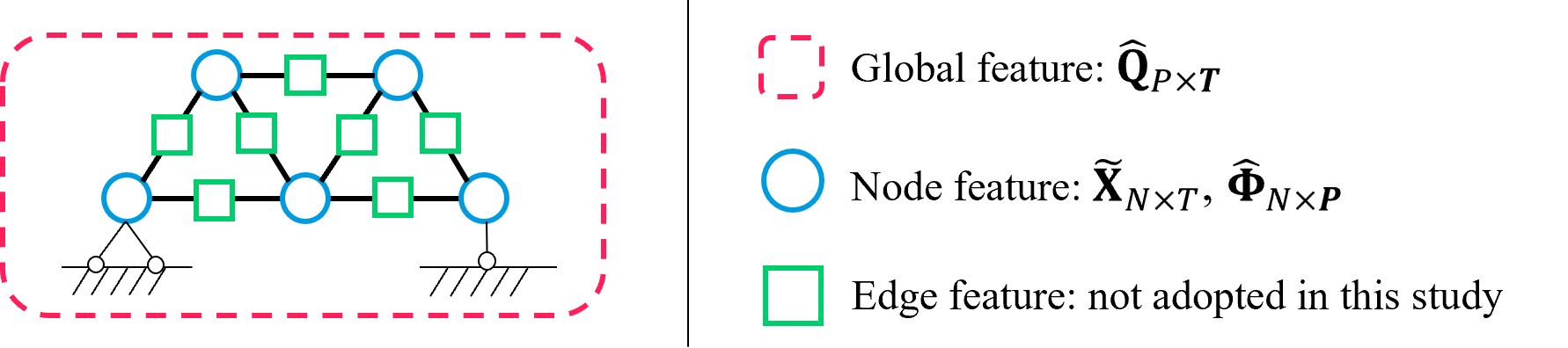}
 \caption{An example of the graph dataset used in this study. A truss structure can be naturally represented by an attributed graph, where node acceleration and mode shapes are node features, and modal responses are graph (global) features.}
    \label{fig:graph}
\end{figure}

As can be seen in Figure \ref{fig:architecture}, three deep learning blocks, which are GraphSAGE, Transformer, and Multi-Layer Perceptron (MLP), are employed to build the entire model. Details about them are given as follows:

1) \textbf{GraphSAGE}. The GraphSAGE model, proposed by Hamilton et al. \cite{hamilton2017inductive}, is a type of GNN, and it serves the key building block in the proposed architecture. Prior research \cite{jian2024using} has demonstrated that GraphSAGE can outperform other GNN variants in the context of PBSHM. Therefore, this study chooses GraphSAGE to process dynamic measurements $\tilde{\mathbf{X}}$, and thereby extract the hidden features $\mathbf{H}$ that contain spatial information from $\tilde{\mathbf{X}}$. As illustrated in Figure \ref{fig:graph}, a structure can be naturally represented by an attributed graph. Based on this graph representation, a GraphSAGE model can be applied by assigning the structure's joint locations as nodes and the structural elements connecting these nodes (e.g., beams, truss bars) as the corresponding edges.

Then, we assign the dynamic measurements at each node $v$, which is denoted as $\tilde{X}_v(t)$, as the initial node feature vector $h_v^{(0)}$ that is fed to GraphSAGE. A GraphSAGE model usually has multiple layers. For each layer $n$, GraphSAGE updates the node features by aggregating information from neighboring nodes. This process is expressed as:

\begin{equation}
\label{eq:SAGE}
h_v^{(n)} = \sigma \left( \mathbf{W}^{(n)} \cdot \max_{u \in \mathcal{N}(v)} \sigma \left( \mathbf{W}_{\text{pool}} \cdot h_u^{(n-1)} + \mathbf{b}_{\text{pool}} \right) \right) + \mathbf{b}^{(n)}
\end{equation}
where $h_v^{(n)}$ is the hidden feature vector of node $v$ at layer $n$. 
$\mathcal{N}(v)$ is the set of neighbors of node $v$. 
$\mathbf{W}^{(n)}$ and $\mathbf{b}^{(n)}$ are the learnable weights and biases of layer $n$.
$\mathbf{W}_{\text{pool}}$ and $\mathbf{b}_{\text{pool}}$ are additional learnable weight and biases applied before the max pooling operation.
$\sigma(\cdot)$ is the non-linear activation function.

After processing through multiple layers, a final hidden feature $\mathbf{H}$ is obtained for each node. Since the operation in Equation \eqref{eq:SAGE} iteratively aggregates the most dominant features from a node’s local neighborhood, the GraphSAGE model is particularly well-suited for capturing spatial dependencies in non-Euclidean domains like a population of structures represented by graphs. While other deep learning models such as CNNs can also learn spatial features in Euclidean settings (like images lying on a grid), they are not inherently designed to model irregular spatial relationships defined by graph structures. Additionally, unlike conventional neural networks with a fixed number of neurons, GraphSAGE can handle graphs with varying numbers of nodes and edges, enabling it to generalize across structures with different geometries and topologies. Consequently, the number of nodes, $N$, can vary between structures, giving GraphSAGE a significant advantage in adapting to a population of structures compared to other deep learning models.

2) \textbf{Set Transformer}. Transformers are advanced deep learning models widely recognized for their ability to effectively extract temporal information from data through their attention mechanism \cite{vaswani2017attention}. However, standard Transformer models are designed for sequential data with a predefined ordering and positional encoding, making them less suited for graph-structured data where input sizes vary and node permutations should not affect the output. To handle these challenges, we adopt the Set Transformer, which is specifically designed to process data with a variable number of nodes while maintaining permutation invariance through the attention mechanism. The details of the Set Transformer are complex and not the focus of this study. Interested readers can refer to \cite{lee2019set} for more information. The set transformer converts the output of GraphSAGE, which is $\mathbf{H}$, into dynamic response $\hat{\mathbf{Q}}$ of SDOF structural modes in the time domain. Since $\hat{\mathbf{Q}}$ is a graph-level (also known as global) feature, this process is named as graph-level task in Figure \ref{fig:architecture}.

3) \textbf{MLP}. Lastly, an MLP, which is the most fundamental deep learning model, is employed to project the hidden feature $\mathbf{H}$ to mode shape vectors $\hat{\mathbf{\Phi}}$ at all nodes. This process is referred to as the node-level task in Figure \ref{fig:architecture} because $\hat{\mathbf{\Phi}}$ is also considered as a node feature.

\subsection{Physics-Informed Loss Function}
The deep learning model proposed in Section \ref{sec:architecture} is designed to identify $\hat{\mathbf{Q}}$ and $\hat{\mathbf{\Phi}}$. However, in modal identification, the ground truth of $\hat{\mathbf{Q}}$ and $\hat{\mathbf{\Phi}}$ is unavailable, posing a significant challenge to train the model in a supervised manner. To tackle this challenge, we adopt a physics-informed learning strategy to train the proposed model in an unsupervised manner. Inspired by prior studies \cite{liu2021machine,bao2024mechanics}, we design the following loss function:

\begin{equation}
\label{eq:loss}
\mathcal{L}=
\lambda_1\cdot\text{MSE}\left(\hat{\mathbf{\Phi}}\cdot\hat{\mathbf{Q}},\tilde{\mathbf{X}}\right)+
\lambda_2\cdot\text{MSE}\left(\mathcal{R}\left(\hat{\mathbf{Q}}\right),\mathbf{I}\right)+
\lambda_3\cdot\text{MSE}\left(\mathcal{R}\left(\left|\text{FFT}\left(\hat{\mathbf{Q}}\right)\right|\right),\mathbf{I}\right)
\end{equation}

Equation \eqref{eq:loss} consists of three terms, with their respective weight coefficients, $\lambda_1$, $\lambda_2$, and $\lambda_3$, that balance the contributions of different loss terms. Each term is informed by a piece of domain knowledge from the field of structural dynamics. The specific details of the applied domain knowledge are as follows:

1) Term 1 applies modal decomposition theory by minimizing mean square errors (MSE) between $\hat{\mathbf{\Phi}}\cdot\hat{\mathbf{Q}}$ and $\tilde{\mathbf{X}}$. This loss term is grounded in the principle that, ideally, the dynamic measurements $\tilde{\mathbf{X}}$ should be equal to the superposition of the dynamic responses of structural vibration modes, represented as $\hat{\mathbf{\Phi}}\cdot\hat{\mathbf{Q}}$.

2) Term 2 enforces the independence property of structural vibration modes in the time-domain. It minimizes the MSE between the correlation coefficient matrix of time-domain dynamic responses of $P$ structural vibration modes, denoted as $\mathcal{R}\left(\hat{\mathbf{Q}}\right)$, and an identify matrix $\mathbf{I}$. Theoretically, the time-domain dynamic response of mode $i$, denoted as $\hat{Q}_i$, is independent of $\hat{Q}_j$ of another mode $j$, implying that their correlation coefficient should be zero. If these correlation coefficients are expressed in matrix form:

\begin{equation}
\label{eq:RQ}
\mathcal{R}\left(\hat{\mathbf{Q}}\right)=\left[\begin{matrix}R_{11}&\cdots&R_{1P}\\\vdots&R_{ij}&\vdots\\R_{P1}&\cdots&R_{PP}\\\end{matrix}\right]_{P \times P}
\end{equation}

where $R_{ij}$ denotes the correlation coefficient between the dynamic response $\hat{Q}_i$ and $\hat{Q}_j$ of vibration mode $i$ and $j$. $R_{ij}$ is computed by:

\begin{equation}
\label{eq:Rij}
R_{ij} = \frac{\sum_{t=1}^{T} (\hat{Q}_{it} - \bar{\hat{Q}}_i)(\hat{Q}_{jt} - \bar{\hat{Q}}_j)}
{\sqrt{\sum_{t=1}^{T} (\hat{Q}_{it} - \bar{\hat{Q}}_i)^2} \sqrt{\sum_{t=1}^{T} (\hat{Q}_{jt} - \bar{\hat{Q}}_j)^2}}
\end{equation}

According to the independence property, all off-diagonal elements in $\mathcal{R}\left(\hat{\mathbf{Q}}\right)$ should be 0, while the diagonal elements should be 1.

3) Term 3 enforces the independent property of structural vibration modes in the frequency-domain, using the amplitude spectrum of fast Fourier transform (FFT) of dynamic responses of modes, denoted as:

\begin{equation}
\left|\mathrm{FFT}\left(\hat{\mathbf{Q}}\right)\right|=\left[\begin{matrix}\left|\mathrm{FFT}\left({\hat{Q}}_1\right)\right|&\left|\mathrm{FFT}\left({\hat{Q}}_2\right)\right|&\cdots&\left|\mathrm{FFT}\left({\hat{Q}}_P\right)\right|\\\end{matrix}\right]^\mathrm{T}
\end{equation}

Since the natural frequencies of different structural vibration modes must be distinct, the corresponding correlation coefficient matrix, $\mathcal{R}\left(\left|\text{FFT}\left(\hat{\mathbf{Q}}\right)\right|\right)$, is supposed to be an identify matrix. The computation of $\mathcal{R}\left(\left|\text{FFT}\left(\hat{\mathbf{Q}}\right)\right|\right)$ follows a similar approach as in Equation \eqref{eq:RQ} and Equation \eqref{eq:Rij}, and is therefore omitted for conciseness.

Finally, it can be seen that employing Equation \eqref{eq:loss} as the loss function for model training eliminates the reliance on ground truth modal parameters, greatly improving the practicality of the proposed modal identification approach.

\subsection{Framework}
\label{sec:framework}
\begin{figure}[!htp] 
   \centering 
   \includegraphics[scale=0.77]{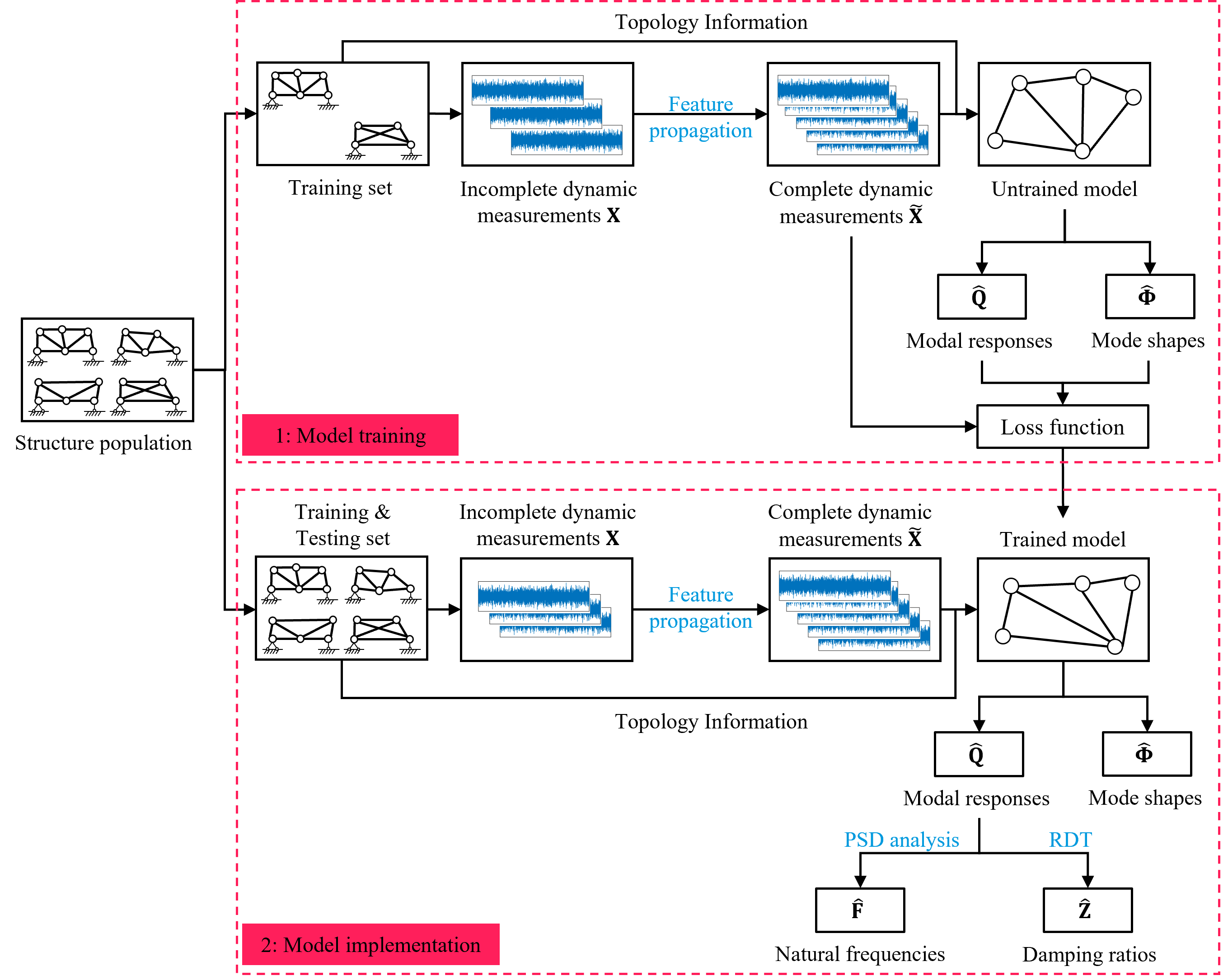}
   \caption{Framework of using the GNN-based model for population-based structural modal identification}
   \label{fig:framework}
\end{figure}

Based on the previously described problem formulation, model architecture, and loss function, the proposed framework for modal decomposition and identification is summarized in Figure \ref{fig:framework}. The figure illustrates that the framework comprises two stages: model training and model implementation.

During the model training stage, a subset of the structural population is used as the training set, which can be either homogeneous or heterogeneous. A \textit{homogeneous population} consists of structures with the same topology but varying load, material, and environmental conditions \cite{bull2021foundations}, while a \textit{heterogeneous population} includes structures with similar yet distinct topologies along with differences in these conditions \cite{gosliga2021foundations}. For each structure in the training set, the Feature Propagation algorithm is first applied to reconstruct complete dynamic measurements from the available incomplete measurements. Subsequently, attributed graphs are constructed using the structural topology, with the complete dynamic measurements assigned as node features. Finally, these graph-structured data are fed into the untrained deep learning model, which outputs modal responses and mode shapes for training the model through the physics-informed loss function.

Once training is complete, the trained model can be directly applied to estimate time-domain modal responses and the corresponding mode shapes for the entire structural population. Subsequently, two widely-accepted modal identification methods, which are Power Spectral Density (PSD) analysis and the Random Decrement Technique (RDT), are employed to extract natural frequencies and damping ratios from the estimated SDOF modal responses.

\section{Numerical Experiments}
In this section, we present the conducted numerical experiments on a heterogeneous population of trusses. The experiments are designed to answer the following research questions: 
\textbf{1)} Can the proposed approach effectively perform modal decomposition and identification across a heterogeneous population of structures? 
\textbf{2)} What are the strengths and limitations of the proposed model compared to existing modal identification methods? 
\textbf{3)} In what ways do the various deep learning blocks in the model architecture and physics-informed terms in the loss function influence overall performance? 

\subsection{Dataset Description}
The extraction of datasets from real-world heterogeneous populations of structures is non-trivial.
While recent initiatives in Asia \citep{Fernando2018Technical} and Europe \citep{limongelli2024bridge} have focused on denser instrumentation of structural populations, such as bridges, these efforts have yet to result in publicly available datasets. Consequently, this study begins its validation through numerical experiments, where a heterogeneous population of simulated simply-supported trusses is generated.

\begin{figure}[!htp] 
   \centering 
   \includegraphics[scale=0.75]{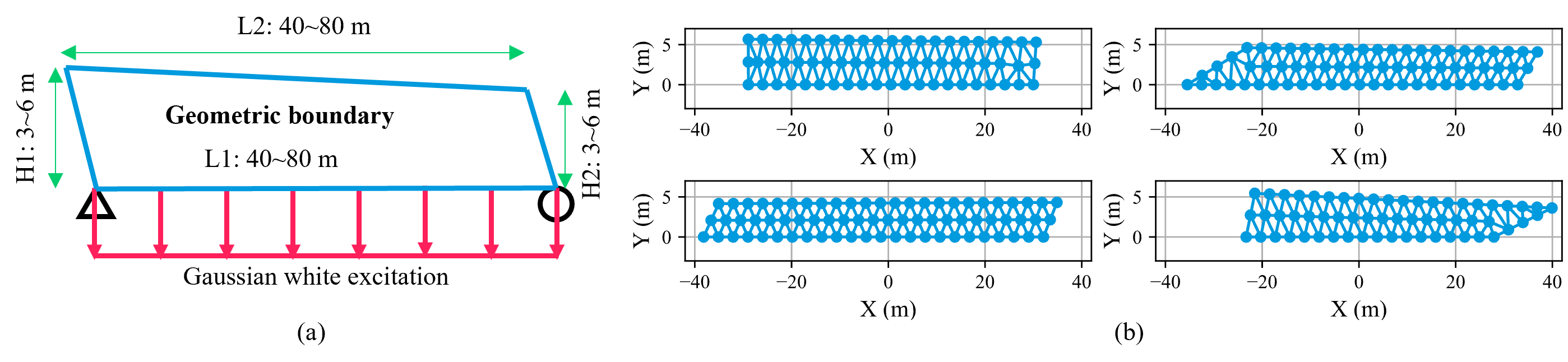}
   \caption{Visualization of the simulated dataset: (a) Geometric configuration designed to approximate a population of simply-supported truss; (b) Representative truss samples from the dataset, displaying only nodes and elements.}
   \label{fig:truss}
\end{figure}

As illustrated in Figure \ref{fig:truss} (a), the truss population is arranged within a trapezoidal boundary, approximating the geometrical configuration of simply-supported beam structures. Each truss is generated using random meshing of the trapezoidal area with Delaunay triangulation \cite{persson2004simple}, which ensures that no point lies within the circumcircle of any triangle, thereby maintaining geometric stability and well-shaped elements. A total of 100 trusses are generated, with 80, 5, and 15 trusses designated for the training, validation, and test datasets, respectively, following standard dataset partitioning practices in deep learning. Examples of the generated trusses are presented in Figure \ref{fig:truss} (b).

\begin{figure}[!htp] 
   \centering 
   \begin{tabular}{cc}
   \includegraphics[scale=0.65]{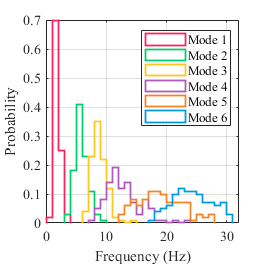} & \includegraphics[scale=0.65]{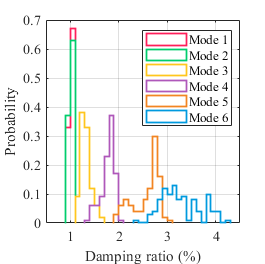} \\
   (a) & (b) \\
   \end{tabular}
    \caption{Probability histograms of the modal properties for the first six modes of the 100 trusses: (a) Natural frequencies; (b) Damping ratios.}
   \label{fig:histograms}
\end{figure}

Based on the generated geometric configurations and boundary conditions (simply-supported type) shown in Figure \ref{fig:truss} (a), corresponding finite element models (FEM) are straightforwardly created using truss elements with realistic structural parameters. The density and area of truss elements are constantly set as 8015 kg/m$^3$ and 0.5 m$^2$, respectively. In order to reflect varying material properties, the Young's modulus of the employed truss elements is set as a random number ranging from 100 GPa to 300 GPa. An eigenvalue analysis is then conducted on all 100 FEMs, generating the reference natural frequencies and mode shapes for model evaluation. The widely-accepted Rayleigh damping model is employed to simulate damping effects. In particular, the modal damping ratio of the first and second vibration modes of every simulated truss is set to 0.01, which allows a calculation of the Rayleigh model coefficients ($\alpha$, $\beta$). Figure \ref{fig:histograms} presents the probability histograms of the natural frequencies and damping ratios for the first six modes across the 100 trusses, highlighting the diversity within the dataset.

\begin{figure}[!htp] 
   \centering 
   \begin{tabular}{cc}
   \includegraphics[scale=0.52]{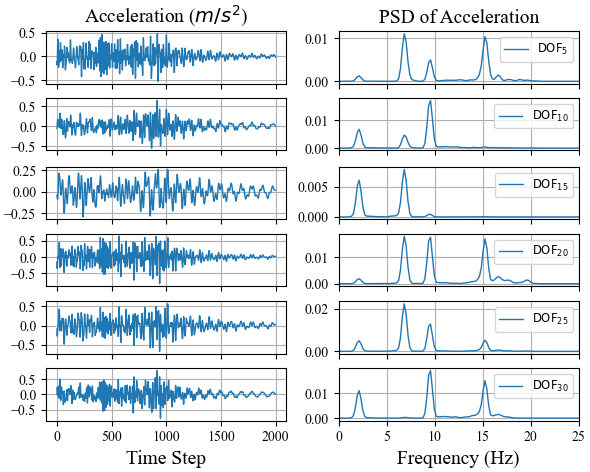} & \includegraphics[scale=0.52]{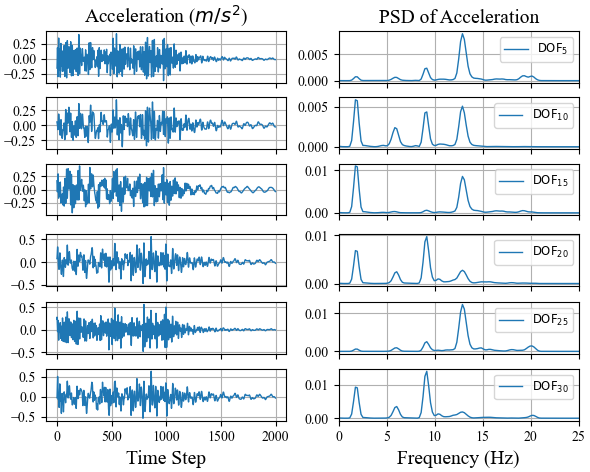} \\
   (a) & (b) \\
   \end{tabular}
    \caption{Filtered vibration acceleration signals for six evenly selected DOFs from two trusses in the simulated truss population, along with their power spectral density (PSD): (a) Truss No. 3, (b) Truss No. 93.}
   \label{fig:measurement_numerical}
\end{figure}

Figure \ref{fig:truss} (a) further indicates the external excitation (Gaussian white noise) that are imposed on the bottom boundary. Furthermore, after the 1000th time step (half the signal length), the Gaussian white noise excitation is removed, allowing the trusses to undergo free decay vibration. This setup is designed to evaluate the proposed approach's performance under nonstationary dynamic measurements, making the scenario more representative of real-world conditions.
Linear time history analyses (Newmark-$\beta$ method) are then performed to obtain the in-plane vertical nodal acceleration (response) time series, of 10 seconds duration, sampled at a time step of 0.005 second (200 Hz). 
To consider the realistic scenario where only a small amount of structural nodes can be monitored, we modify the generated dataset by evenly removing time series data from 82\% of the nodes in each truss. The reason for evenly selecting nodes for removal, rather than randomly, is that sensors are typically evenly distributed on monitored structures in practice. The complete dynamic measurements will be reconstructed by the Feature Propagation algorithm and max-normalized (scaled to a maximum value of 1) to facilitate the use of the proposed deep learning model. 
In the numerical experiments, we focus on identifying the properties of the first 4 structural vibration modes. Among the 100 simulated trusses, we observe that the maximum frequency of the fourth mode is around 20 Hz, as shown in Figure \ref{fig:histograms} (a). Therefore, before feeding the simulated acceleration time series into the deep learning model, we apply a 20-Hz low-pass filter to remove higher-frequency components.
Figure \ref{fig:measurement_numerical} shows the filtered vibration acceleration signals of six evenly selected DOFs from two trusses within the simulated truss population. As illustrated in Figure \ref{fig:measurement_numerical}, the time-domain acceleration signals exhibit nonstationary behavior. The first half of the signals correspond to forced vibration induced by Gaussian white noise excitation, while the second half represents free decay vibration after the excitation is removed. The frequency-domain PSD confirms that frequency components above 20 Hz have been effectively filtered out.

\subsection{Implementation Details}
\label{sec:implementation}
Our deep learning model is trained and implemented using PyTorch 2.0.1 \cite{paszke2019pytorch}, CUDA 11.7, and DGL 1.1.1 \cite{wang2019deep}. The Adam optimizer \cite{kingma2014adam} is used for training with default settings, including a learning rate of 0.0003, a first momentum decay parameter of 0.9, and a second momentum decay parameter of 0.999. To accelerate the batch training process, we utilize an NVIDIA GeForce RTX 3060. The model is trained with a batch size of 64 for 5,000 epochs. Both MLPs and GNNs in our study consist of three layers, with each layer having a dimension of 128. In the loss function, the coefficients $\lambda_1$ (modal decomposition theory), $\lambda_2$ (independent property of modal responses in the time domain), and $\lambda_3$ (independent property of modal responses in the frequency domain) are set to 10, 1, and 1, respectively. Additionally, since this study focuses on modal identification of the first 4 modes, we set the hyperparameter $P$ of our model to 7, slightly exceeding 4, to accommodate noise and redundant modes with frequencies below 20 Hz.

More implementation details on top of the settings introduced above are available in our GitHub repository \cite{Jian2024}, which will be made publicly accessible upon publication of this work. Furthermore, it is important to note that this study aims to demonstrate the feasibility of the proposed approach rather than presenting a fully optimized model with the best performance. While we performed some hyper-parameter tuning, it is not exhaustive, as our primary goal is to validate the functionality of the network without requiring extensive optimization. Interested readers can further fine-tune the hyper-parameters using our publicly available code.

\subsection{Experimental Results}
\begin{figure}[!htp] 
   \centering 
   \includegraphics[scale=0.7]{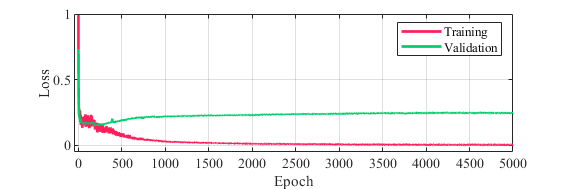} 
    \caption{Training and validation loss curves from the training process.}
   \label{fig:loss_numerical}
\end{figure}

As outlined in Section \ref{sec:framework}, the first step in applying the proposed approach for modal decomposition and identification is model training. Figure \ref{fig:loss_numerical} presents the loss curves from the training process, where 80 trusses are used for training and 10 trusses for validation. The loss is computed using Equation \eqref{eq:loss}. 



Figure \ref{fig:loss_numerical} allows for the following observations: \textbf{1)} Both training and validation losses decrease significantly at the beginning and eventually converge, indicating an overall effective training process. \textbf{2)} The training loss is lower than the validation loss, which is expected, as the validation loss is not involved in backpropagation for updating the learnable model parameters. \textbf{3)} The validation loss converges faster than the training loss. However, the training process is not halted at that point, which is common practice in deep learning. In this study, the loss function is entirely physics-informed and no ground-truth modal properties are available, meaning that the training loss itself reflects the model’s capability to capture physically consistent modal behaviors. Therefore, the training is continued until both losses reach convergence.

\begin{figure}[!htp] 
   \centering 
   \begin{tabular}{c}
   \includegraphics[scale=0.52]{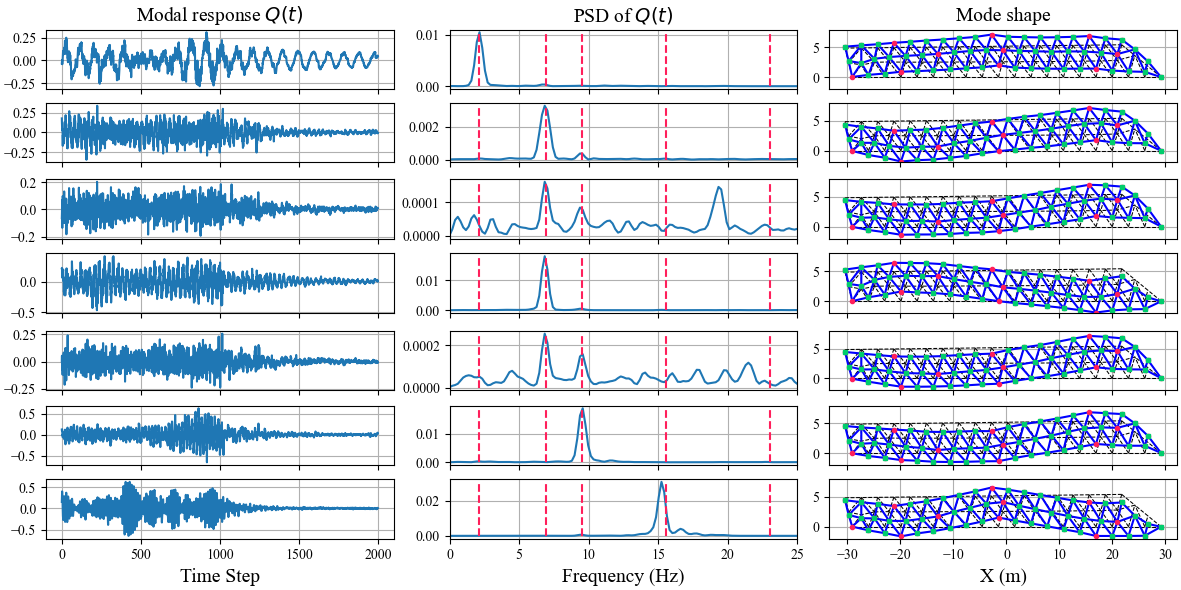} \\ 
   (a) \\
   \includegraphics[scale=0.52]{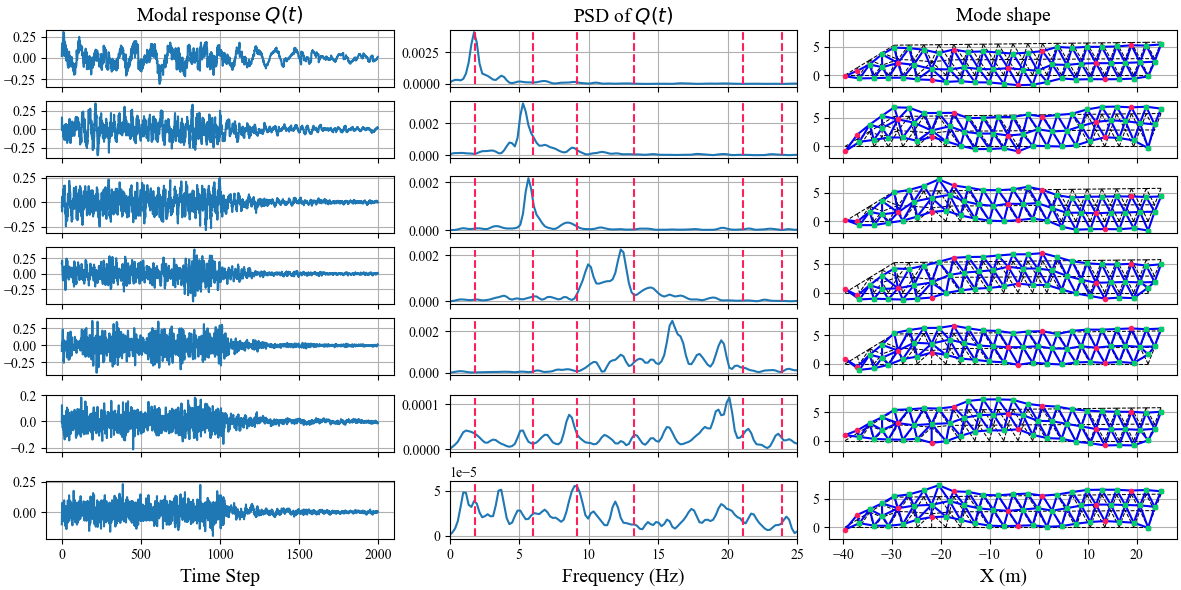} \\
   (b) \\
   \end{tabular}
    \caption{Modal decomposition results for two representative trusses. Each row corresponds to a decomposed mode. The first column displays the time-domain dynamic responses of the decomposed modes. The second column presents the PSD of the dynamic modal responses, with the true mode frequencies, obtained via eigenvalue analysis, indicated by red dashed lines. The third column visualizes the corresponding mode shapes, where the truss topology is represented by black dashed lines, structural nodes with known measurements are marked in red, and structural nodes without measurements are marked in green. These results correspond to: (a) Truss No. 3, whose acceleration signals are shown in Figure \ref{fig:measurement_numerical} (a) and is included in the training set; and (b) Truss No. 93, whose acceleration signals are shown in Figure \ref{fig:measurement_numerical} (b) and was not used during training.}
   \label{fig:decomposition}
\end{figure}

Upon completing the training, the trained model is utilized for modal decomposition and identification across all 100 trusses in the simulated truss population. The complete acceleration measurements, reconstructed using the FP algorithm, are fed into the trained model, generating the decomposed dynamic responses of the SDOF modes along with their corresponding mode shapes. Figure \ref{fig:decomposition} presents the modal decomposition results for the two trusses shown in Figure \ref{fig:measurement_numerical}, among which Figure \ref{fig:decomposition} (a) illustrates a truss that is included in the training set, while Figure \ref{fig:decomposition} (b) depicts a truss that is not used during training.

Figure \ref{fig:decomposition} presents the following key observations: \textbf{1)} As shown in Figure \ref{fig:decomposition} (a), the trained model demonstrates effective performance on a truss that is included in the training set. It is important to note that the modal properties of the trusses in the training set remain unknown, meaning that good performance on the training set, which typically implies overfitting, is not a concern in this study. 
The acceleration measurements of this MDOF truss are effectively decomposed into the dynamic responses of SDOF modes. In the time domain, the waveforms of these modal responses exhibit nonstationary characteristics, with the first half representing forced vibration and the second half showing free decay vibration. In the frequency domain, most PSD plots in the second column of Figure \ref{fig:decomposition} display a single dominant peak, which aligns well with the true modal frequencies (indicated by red dashed lines), confirming the accuracy of the extracted modal frequencies.
Lastly, the corresponding identified mode shapes also appear physically reasonable. \textbf{2)} Some PSD plots in Figure \ref{fig:decomposition} (a)—notably in the third and fifth rows of the PSD column—exhibit multiple peaks, which do not correspond to actual structural vibration modes. However, these spurious modes can be easily distinguished, not only because they exhibit multiple peaks, but also due to their significantly lower magnitudes compared to the true modes. For instance, the third and fifth PSD plots in Figure \ref{fig:decomposition} (a) have peak magnitudes of approximately 0.0001 and 0.0002, whereas the true structural modes have magnitudes at least an order of magnitude larger. \textbf{3)} Figure \ref{fig:decomposition} (b) demonstrates that the trained model is capable of performing modal decomposition on an unseen truss that is not included in the training set. The first and second modes are partially separated, indicating some level of generalization. However, the overall decomposition quality is suboptimal. Specifically, the PSD of the decomposed modal responses does not consistently exhibit a single dominant peak, suggesting imperfect mode separation. Moreover, the identified mode shapes appear irregular and lack smoothness, which is inconsistent with expectations from structural analysis theory.

Next, we evaluate the proposed approach within the population-based SHM framework. The RDT and PSD analysis are applied to the dynamic responses of the decomposed SDOF modes to extract damping ratios and natural frequencies of the first four modes across all 100 trusses. The accuracy of these identified parameters is assessed by comparing them with the true frequencies and damping ratios obtained from FEM modal analysis. Additionally, the Modal Assurance Criterion (MAC), defined by Equation \eqref{eq:MAC}, is computed to evaluate the accuracy of the identified mode shapes. A MAC value close to 1 indicates a high degree of accuracy in mode shape identification.

\begin{equation}
    \label{eq:MAC}
    \text{MAC}(\hat{\phi}, \phi) = \frac{ \left ( \hat{\phi}^\mathrm{T} \phi \right )^2}{(\hat{\phi}^\mathrm{T} \hat{\phi})(\phi^\mathrm{T} \phi)}
\end{equation}
where $\hat{\phi}$ and $\phi$ are vectors of the identified and true mode shapes, respectively, at all structural nodes.

Table \ref{tab:statistics} summarizes the statistical results of modal identification across the entire truss population, leading to the following key observations: \textbf{1)} The trained model performs well on Trusses 1–80, which are included in the training set. This demonstrates that the model can effectively achieve modal decomposition and identification for a population of structures. \textbf{2)} In contrast, modal identification results for Trusses 81–100, which are not included in the training set, are less satisfactory, even though modes that are not effectively separated are excluded from the statistical analysis. This suggests that the generalization ability of the model is limited. However, this limitation does not undermine the effectiveness of the proposed approach, as the physics-informed training process is essentially the process of conducting modal decomposition and identification. Moreover, existing studies on mechanics-informed deep learning for modal identification \cite{liu2021machine, bao2024mechanics} have not reported investigations into generalization ability, making this study a valuable attempt to extend a trained deep learning model to a population of structures.

\begin{table}[!htp]
\centering
\caption{Statistics of modal identification results for the entire truss population}
\label{tab:statistics}
\begin{tabular}{cccccccc}
\toprule
     &        & \multicolumn{3}{c}{\begin{tabular}[c]{@{}c@{}}Truss 1-80 \end{tabular}} & \multicolumn{3}{c}{\begin{tabular}[c]{@{}c@{}} Truss 81-100\end{tabular}} \\ \midrule
     &        & Mean & Median  & Std & Mean & Median  & Std          \\
\midrule
MAC  & mode 1 & \textbf{0.968}& 0.976& 0.028 & \textbf{0.512}   & 0.525     & 0.323  \\
     & mode 2 & \textbf{0.869}& 0.933& 0.182 & \textbf{0.541}   & 0.566     & 0.282  \\ 
     & mode 3 & \textbf{0.802}& 0.863& 0.190 & \textbf{0.179}   & 0.103     & 0.181  \\
     & mode 4 & \textbf{0.835}& 0.883& 0.124 & \textbf{0.370}   & 0.329     & 0.298  \\
\midrule
Frequency & mode 1 & 1.063& 0.884& 2.960 & -0.043  & 1.001     & 8.915  \\
errors     & mode 2 & -0.880  & -0.404  & 2.927 & -1.452  & -1.504    & 6.957  \\
(\%)     & mode 3 & 0.428& -0.232  & 3.590 & 6.307   & 7.747     & 4.830  \\
     & mode 4 & -1.615  & -1.359  & 2.303 & 0.186   & 1.884     & 8.407  \\
\midrule
Damping & mode 1 & 54.641  & 34.170  & 81.458& 124.658 & 146.105   & 66.302 \\
ratio     & mode 2 & 77.242  & 84.357  & 76.522& 134.458 & 129.784   & 31.401 \\
errors     & mode 3 & 51.669  & 49.888  & 52.869& 62.864  & 72.077    & 32.251 \\
(\%)  & mode 4 & -12.248 & -13.987 & 33.007& 41.638  & 39.227    & 20.708  \\    
\bottomrule
\end{tabular}
\end{table}

\subsection{Comparison against Classic Modal Identification Methods }
To comprehensively evaluate our model in modal identification, we compare its performance against two of the most widely used and representative modal identification methods: Enhanced Frequency Domain Decomposition (EFDD) \cite{brincker2001damping}, which operates in the frequency domain, and Stochastic Subspace Identification (SSI) \cite{peeters1999reference}, which operates in the time domain. While SSI is fully automatic, the original EFDD method requires a manual peak-picking process to select the modes for identification. However, given the large truss population in our test set, manual peak selection would be impractical for population-based modal identification. To address this challenge, we employ the Automated EFDD algorithm proposed by \cite{cheynet2017damping}, which enables automatic identification of mode shapes, damping ratios, and natural frequencies for the first four vibration modes of each simulated truss. It is also worth mentioning that, for mode shape identification using Automated EFDD and SSI, both methods initially extract mode shapes from the 18\% available dynamic measurements. Afterward, interpolation and extrapolation are applied to extend the identified mode shapes to all nodes along the beam, based on structural node coordinates.

Table \ref{tab:statistics_comparison} presents the statistical results of modal identification for our approach, EFDD, and SSI. This comparison is conducted only on Trusses 1–80, as the performance of our approach on unseen trusses is limited. The statistics in Table \ref{tab:statistics} lead to the following key observations: \textbf{1)} Our approach achieves comparable accuracy to EFDD and SSI in identifying natural frequencies, damping ratios, and low-order mode shapes. \textbf{2)} Our approach outperforms EFDD and SSI in identifying high-order mode shapes. This is primarily due to its ability to handle incomplete measurements, where dynamic measurements at 82\% of nodes are missing. Our approach leverages structural topology by applying the Feature Propagation algorithm to reconstruct complete dynamic measurements before modal identification. In contrast, EFDD and SSI do not incorporate structural topology. They identify mode shapes only at nodes with available measurements and subsequently rely on interpolation and extrapolation to estimate mode shapes at unmeasured nodes.

\begin{table}[!htp]
\centering
\caption{Statistics of modal identification results for truss 1-80: comparison against classic modal identification methods}
\label{tab:statistics_comparison}
\begin{tabular}{ccccccccccc}
\hline
          &        & \multicolumn{3}{c}{Our approach}    & \multicolumn{3}{c}{EFDD}   & \multicolumn{3}{c}{SSI} \\ \hline
          &        & Mean    & Median  & Std    & Mean    & Median  & Std    & Mean    & Median  & Std     \\ \hline
MAC       & mode 1 & 0.968   & 0.976   & 0.028  & 0.988   & 0.994   & 0.017  & 0.944   & 0.991   & 0.181   \\
          & mode 2 & 0.869   & 0.933   & 0.182  & 0.848   & 0.909   & 0.184  & 0.745   & 0.891   & 0.294   \\
          & mode 3 & \textbf{0.802}   & 0.863   & 0.190  & \textbf{0.740}   & 0.828   & 0.248  & \textbf{0.699}   & 0.798   & 0.267   \\
          & mode 4 & \textbf{0.835}   & 0.883   & 0.124  & \textbf{0.592}   & 0.654   & 0.253  & \textbf{0.529 }  & 0.617   & 0.297   \\ \hline
Frequency & mode 1 & 1.063   & 0.884   & 2.960  & -0.671  & -0.222  & 11.208 & 0.233   & 0.160   & 1.542   \\
errors    & mode 2 & -0.880  & -0.404  & 2.927  & -0.256  & -0.405  & 4.298  & 0.878   & -0.226  & 7.142   \\
(\%)      & mode 3 & 0.428   & -0.232  & 3.590  & 1.248   & -0.240  & 7.980  & 1.122   & -0.479  & 7.638   \\
          & mode 4 & -1.615  & -1.359  & 2.303  & -2.507  & -2.198  & 3.459  & -1.808  & -0.928  & 5.846   \\ \hline
Zeta      & mode 1 & 54.641  & 34.170  & 81.458 & -55.305 & -54.231 & 10.733 & 164.318 & 92.756  & 233.524 \\
ratio     & mode 2 & 77.242  & 84.357  & 76.522 & -75.464 & -76.422 & 4.709  & 11.328  & -1.448  & 66.537  \\
errors    & mode 3 & 51.669  & 49.888  & 52.869 & -83.452 & -84.012 & 3.303  & -18.038 & -34.341 & 53.157  \\
(\%)      & mode 4 & -12.248 & -13.987 & 33.007 & -89.360 & -89.935 & 2.687  & -32.130 & -46.510 & 49.886  \\ \hline
\end{tabular}
\end{table}

\begin{figure}[!htp] 
   \centering 
   \includegraphics[scale=0.54]{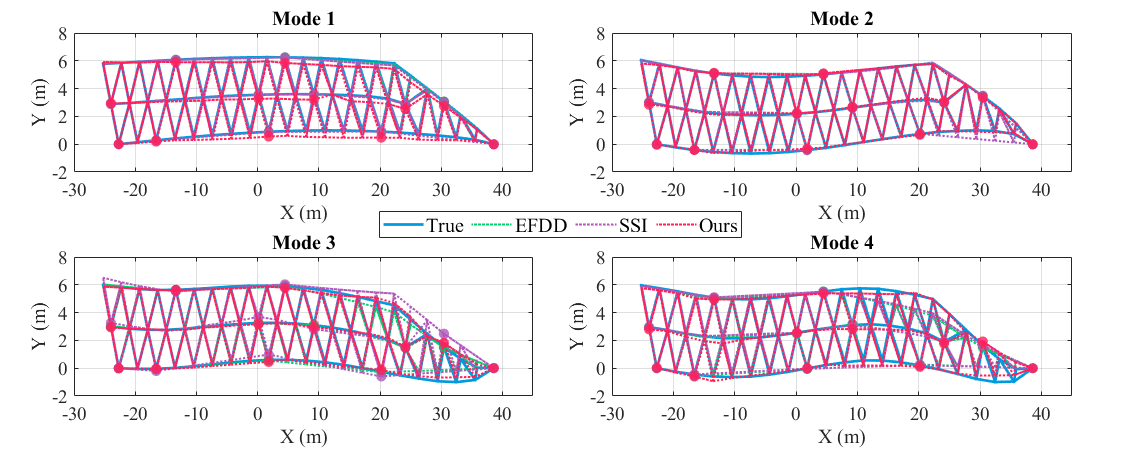} 
    \caption{True and identified mode shapes of a representative truss, in which structural nodes with measurements are highlighted as solid dots.}
   \label{fig:modeshape}
\end{figure}

To further demonstrate the superior performance of our approach in identifying high-order mode shapes, Figure \ref{fig:modeshape} presents the first four true and identified mode shapes for a representative truss example. Notably, in the third and fourth modes, the mode shapes identified by our approach exhibit a significantly better alignment with the true mode shapes compared to those obtained using EFDD and SSI. This observation is further supported by the computed MAC values. For our approach, the MAC values for the four identified mode shapes are 0.96657, 0.97174, 0.94341, and 0.77402, respectively. In contrast, EFDD yields 0.99503, 0.90982, 0.68174, and 0.28856, while SSI produces 0.99339, 0.91109, 0.37682, and 0.28488, demonstrating the notable advantage of our approach in accurately capturing high-order mode shapes.

In addition to accuracy, we also compare the computational efficiency of the three approaches. The total time required for modal identification of 80 trusses using EFDD, SSI, and our approach is 18.016, 138.061, and 1682.599 seconds, respectively. Our approach is significantly slower than EFDD and SSI, primarily due to the deep learning model training, which accounts for 1677.801 seconds, even though the model inference takes only 4.797 seconds. However, real-time modal decomposition and identification are generally not required, making the longer computation time acceptable. Furthermore, this computational expense is compensated by our approach’s unique capability to separate dynamic responses of individual structural vibration modes, and that is a feature absent in EFDD and SSI.

At last, for the reproducibility of this study, the MATLAB implementations of both the Automated EFDD and SSI algorithms will be made publicly available in a dedicated GitHub repository \cite{Jian2024} upon the publication of this work. Interested readers can also find detailed setting of these two algorithms in the codes, which are omitted in this paper for conciseness.

\subsection{Ablation Study}
To validate the effectiveness of our model architecture and physics-informed loss function, we conduct an ablation study by comparing the originally proposed approach with the following modified variants:

\textbf{1) No GNN}: We remove the GraphSAGE block in the model architecture (Figure \ref{fig:architecture}). Then we use the standard Transformer and MLP to directly process the dynamic measurements to obtain dynamic responses of modes and corresponding mode shapes. However, removing the GNN eliminates the model's ability to handle graph-structured data with varying node counts $N$, as both the standard Transformer and MLP require a fixed input size. As a consequence, we have to constrain all trusses to only $P$ selected nodes ($P \ll N$) to ensure compatibility, though the model's capacity is reduced.

\textbf{2) No Transformer}: We replace the Set Transformer block with the Set Long Short-Term Memory (LSTM) in the model architecture. LSTM is another type of deep learning model that has been broadly used to handle time series data \cite{yu2019LSTM}. However, standard LSTMs require fixed input dimensionality at each time step and are not inherently designed to handle varying numbers of input nodes, making them unsuitable for this study, where different structures produce dynamic measurements of varying size. To address this limitation, we adopt Set LSTM, which is specifically designed to process unordered sets with varying sizes and is invariant to input permutations. \cite{vinyals2016ordermatterssequencesequence}. 

\textbf{3) No mode independence}: In the loss function expressed in Equation \eqref{eq:loss}, we remove the second and third terms that enforce the independent property of structural vibration modes.  In this case, the model is trained solely based on the modal decomposition term, expressed as $\text{MSE}\left(\hat{\mathbf{\Phi}}\cdot\hat{\mathbf{Q}},\tilde{\mathbf{X}}\right)$.

\begin{figure}[!htp] 
   \centering 
   \begin{tabular}{ccc}
   \includegraphics[width=5.2cm]{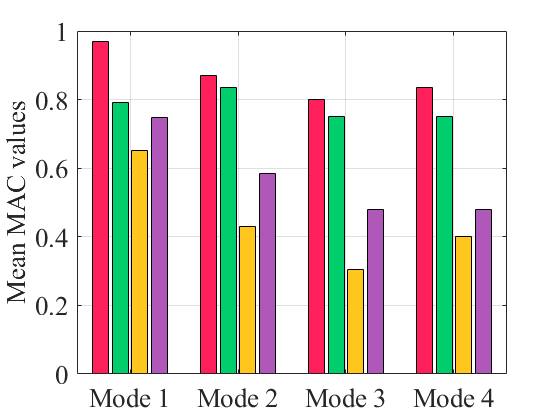} & \includegraphics[width=5.2cm]{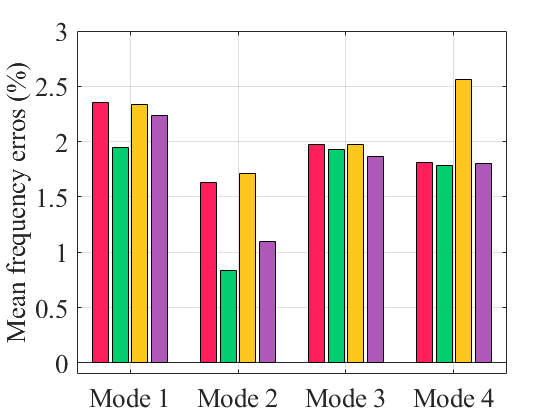} & \includegraphics[width=5.2cm]{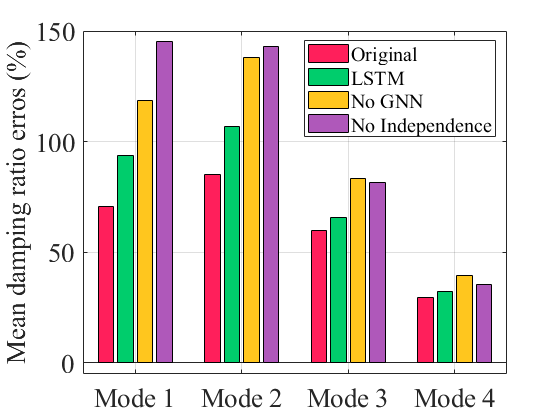} \\
   (a) & (b) & (c) \\
   \end{tabular}
    \caption{Modal identification results of the originally proposed approach and its variations: (a) Mean MAC values of the identified mode shapes, (b) Mean absolute percentage errors in identified frequencies, and (c) Mean absolute percentage errors in identified damping ratios.}
   \label{fig:ablation}
\end{figure}

Figure \ref{fig:ablation} shows the modal identification results of the originally proposed approach and its variations. It provides the following findings: \textbf{1)} The originally proposed approach achieves the best overall performance, exhibiting the highest mean MAC values, the lowest mean errors in identified damping ratios, and comparable mean errors in identified frequencies relative to other variants. These results confirm the superiority of the proposed approach. \textbf{2)} In Figure \ref{fig:ablation} (a), the ‘No GNN’ variant performs worse than any approach that incorporates GNNs, highlighting the effectiveness of GNNs in handling graph-structured data, such as PBSHM datasets. \textbf{3)} The proposed approach also outperforms the ‘LSTM’ variant, which can be attributed to the fact that the study involves relatively long time series (2,000 data points). Transformers, leveraging their attention mechanism, are generally more effective than LSTMs in processing long time series, explaining this performance difference \cite{reza2022multi}. \textbf{4)} In Figure \ref{fig:ablation} (c), the variant that omits the mode independence constraint yields the worst performance, which is expected. Without explicitly enforcing the independent property of structural modes, the dynamic responses of SDOF modes cannot be effectively separated, leading to inferior results.

\section{Laboratory Experiments}
This section presents our laboratory experiments conducted on a scale cable-stayed bridge. The experiments aim to address the following research questions: \textbf{1)} Can the proposed approach effectively carry out modal decomposition and identification on real-world structural dynamic measurements, particularly when the structure is subjected to different types of loading, which can be considered, to some extent, as a homogeneous population of structures? \textbf{2)} How does the performance of the proposed approach compare to classical modal identification methods when applied to real-world datasets?

\subsection{Dataset Description and Implementation Details}
\begin{figure}[!htp] 
   \centering 
   \begin{tabular}{c}
   \includegraphics[scale=0.9]{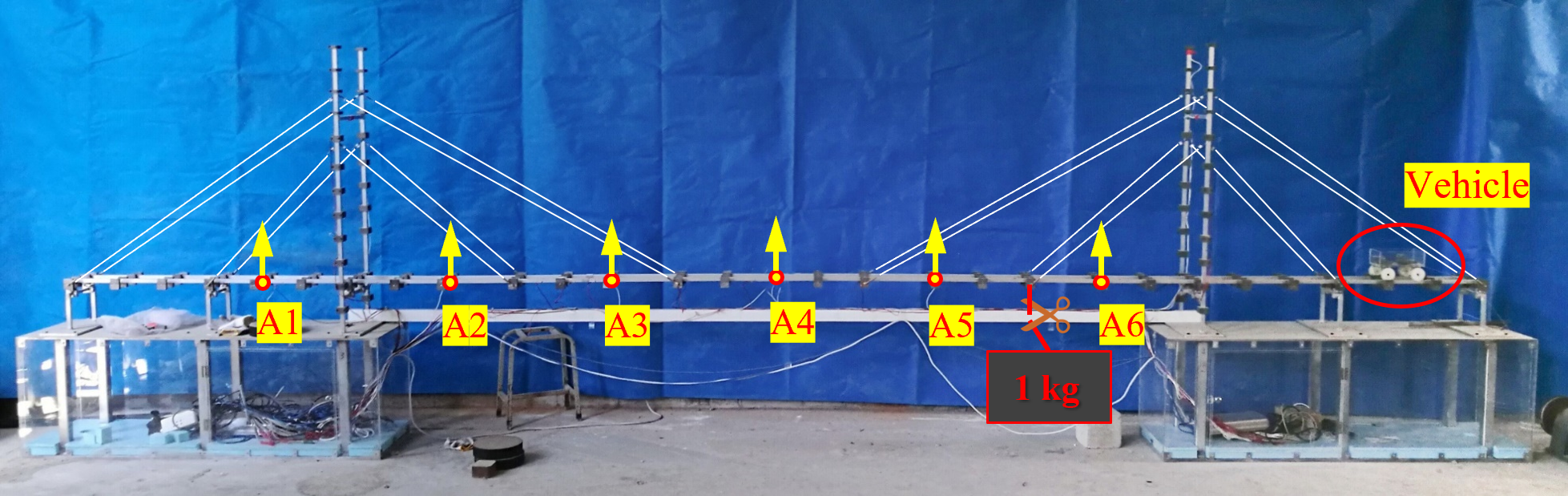} \\ 
   (a) \\
   \includegraphics[scale=0.9]{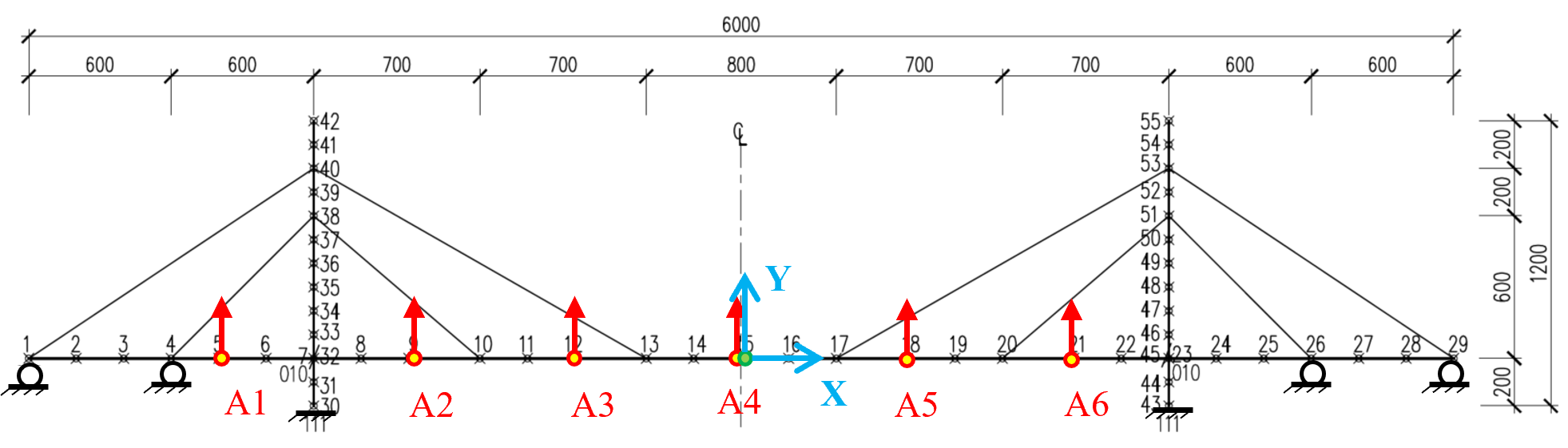} \\
   (b) \\
   \includegraphics[scale=0.57]{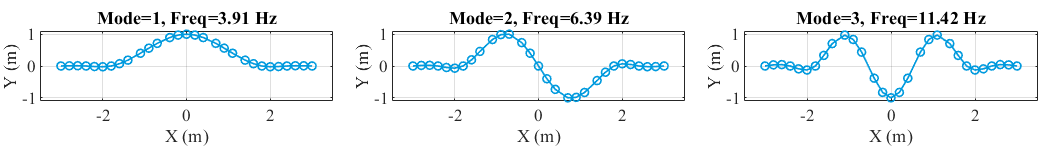} \\
   (c) \\
   \end{tabular}
    \caption{Scale model of a cable-stayed bridge: (a) Photo; (b) Diagram of the finite element model (unit: mm). The six deployed accelerometers are labeled from A1 to A6, with arrows indicating the sensing direction; (c) Frequencies and mode shapes of the first three vertical modes obtained from the eigenvalue analysis of the finite element model.}
   \label{fig:bridge}
\end{figure}

Given the difficulty of gathering data from a heterogeneous population of real-world structures, the laboratory experiments are designed to simulate a partially homogeneous structural population by performing tests on the same structure under different loading conditions. As shown in Figure \ref{fig:bridge} (a), a scaled model of a cable-stayed bridge is constructed, consisting of a 6-meter continuous beam, two towers, and 16 cables. The beam and towers are made of aluminum alloy, with additional metal weights attached to replicate the dynamic properties of a real cable-stayed bridge. To capture its vertical vibration response, six Micro-Electro-Mechanical System (MEMS) accelerometers (A1 to A6) are installed on the beam, recording acceleration data at a sampling frequency of 200 Hz. Two excitation methods are designed to induce vibrations in the bridge model: \textbf{1) Pull-and-Release Method}: As shown at the bottom of Figure \ref{fig:bridge} (a), a 1 kg iron weight is suspended from the beam using a wire. Once both the bridge model and the weight are stationary, the wire is abruptly cut, triggering damped free vibration. The weight is placed precisely at the lateral center of the beam to minimize out-of-plane vibrations such as torsion. \textbf{2) Vehicle Loads}: As depicted on the right side of Figure \ref{fig:bridge} (a), a 2 kg vehicle model is deployed on the beam, using it as a track for movement. The vehicle is connected to a motor via a wire, and the motor rotated at a constant speed, propelling the vehicle along the beam at approximately 0.2 to 0.5 m/s. This motion induced forced vibrations in the bridge model, simulating real-world loading conditions.

As illustrated in Figure \ref{fig:bridge} (b), a two-dimensional finite element model (FEM) of the scaled bridge is created in MATLAB to analyze its modal properties. An eigenvalue analysis is conducted on the FEM model, and the first three vertical mode shapes of the beam, along with their corresponding frequencies, are presented in Figure \ref{fig:bridge} (c).

\begin{figure}[!htp] 
   \centering 
   \begin{tabular}{cc}
   \includegraphics[scale=0.52]{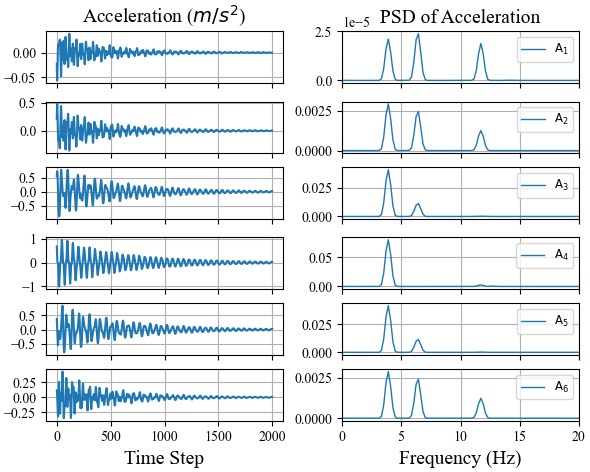} & \includegraphics[scale=0.52]{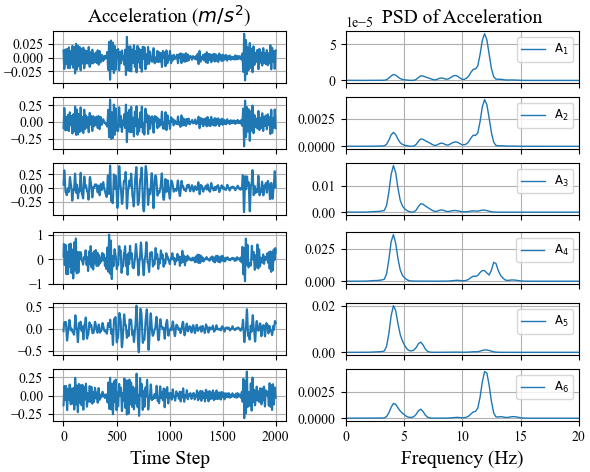} \\
   (a) & (b) \\
   \end{tabular}
    \caption{Filtered vibration acceleration signals from six accelerometers in two representative tests, along with their power spectral density (PSD): (a) Test 1, pull-and-release, (b) Test 7, vehicle-excited.}
   \label{fig:measurement_laboratory}
\end{figure}

A total of 10 tests are conducted on the scaled bridge model, comprising 5 pull-and-release tests and 5 vehicle-excited tests. From each test, 10-second signal segments (2000 time steps) are extracted from the full time-history signals. Using these segments, the feature propagation algorithm is applied to reconstruct acceleration signals at all structural nodes on the beam, followed by max-normalization to optimize the data for deep learning. Data from four pull-and-release tests and four vehicle-excited tests are used as the training set, while the remaining one pull-and-release test and one vehicle-excited test data are reserved for validation and testing. Since this study focuses on identifying the first three vertical modes of the bridge’s beam, a 15 Hz low-pass filter was applied to the acceleration measurements, considering that the third vertical mode's frequency is 11.42 Hz. The filtered acceleration signals from all six accelerometers for two representative tests are shown in Figure \ref{fig:measurement_laboratory}.

The implementation details of our deep learning model for the laboratory experiments largely mirror those used in the numerical experiments, as described in Section \ref{sec:implementation}. Therefore, to avoid redundancy, we do not repeat them here. The only two differences are: 1) the learning rate is set to 0.0001, and 2) the number of decomposed modes ($P$) is set to 5, slightly exceeding the three target modes to accommodate measurement noise and redundant modes.

\subsection{Experimental Results}
\begin{figure}[!htp] 
   \centering 
   \includegraphics[scale=0.7]{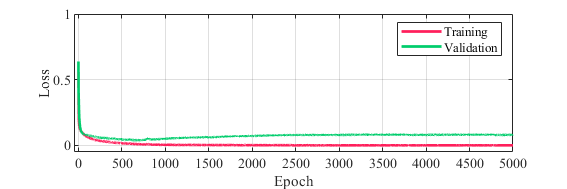} 
    \caption{Training and validation loss curves from the training process.}
   \label{fig:loss_laboratory}
\end{figure}

Figure \ref{fig:loss_laboratory} illustrates the loss curves from the training process, where data from eight tests were used for training and the remaining two tests for validation. Both training and validation losses exhibit a sharp decline initially and eventually converge, demonstrating the effectiveness of the training process.

\begin{figure}[!htp] 
   \centering 
   \begin{tabular}{c}
   \includegraphics[scale=0.52]{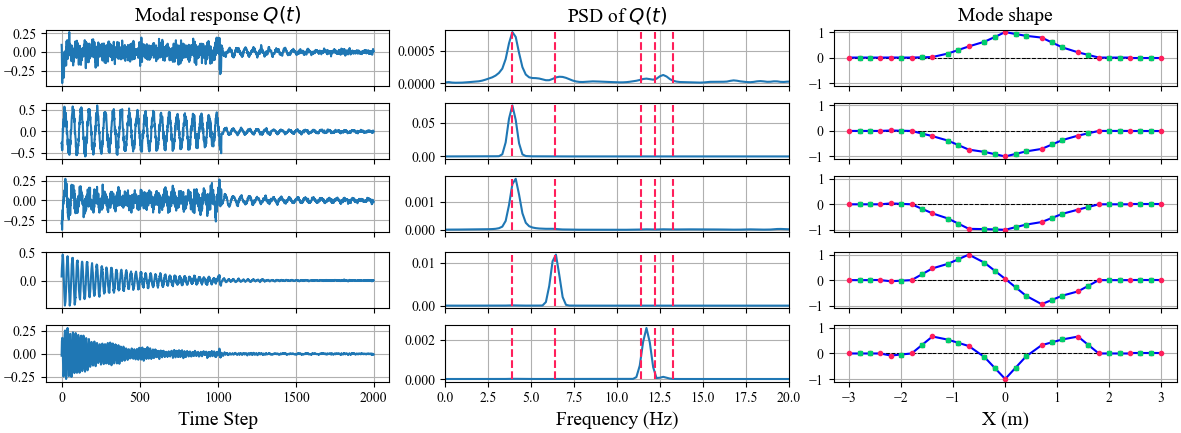} \\ 
   (a) \\
   \includegraphics[scale=0.52]{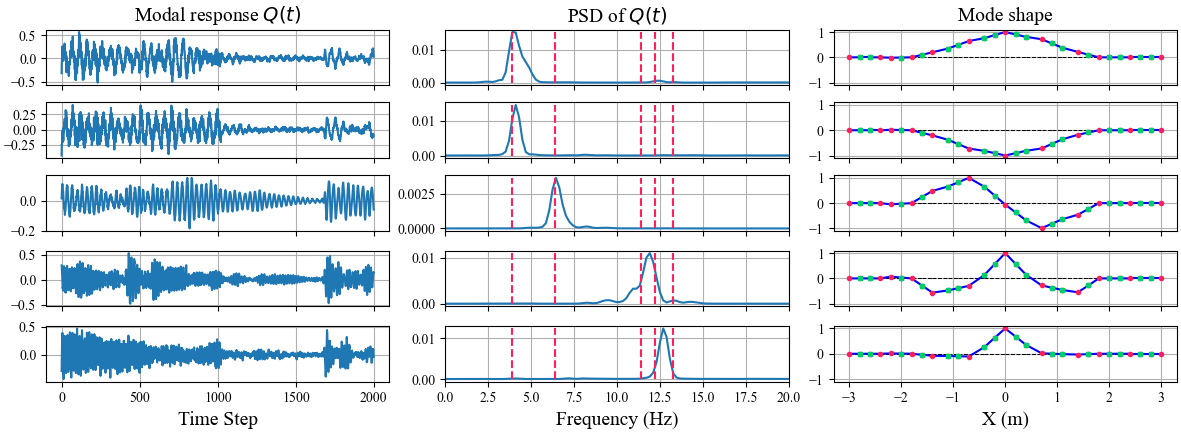} \\
   (b) \\
   \end{tabular}
    \caption{Modal decomposition results for two representative tests. Each row corresponds to a decomposed mode. The first column displays the time-domain dynamic responses of the decomposed modes. The second column presents the PSD of the dynamic modal responses, with the reference frequencies, obtained via eigenvalue analysis of FEMs, indicated by red dashed lines. The third column visualizes the corresponding mode shapes, where the beam topology is represented by black dashed lines, structural nodes with known measurements are marked in red, and structural nodes without measurements are marked in green. These results correspond to: (a) Test No. 1, which is excited by the pull-and-release action, and its acceleration signals are shown in Figure \ref{fig:measurement_laboratory} (a); and (b) Test No. 7, which is excited by vehicle loads, and its acceleration signals are shown in Figure \ref{fig:measurement_laboratory} (b).}
   \label{fig:decomposition_laboratory}
\end{figure}

Next, we apply the trained model for modal decomposition and identification across all 10 tests. Figure \ref{fig:decomposition_laboratory} presents the decomposition results, demonstrating that the model effectively separates the acceleration measurements of the MDOF bridge structure into the dynamic responses of SDOF modes, regardless of the excitation method used. The PSD plots of the decomposed modal responses confirm that their frequencies align well with those obtained from finite element modal analysis, and the extracted mode shapes closely match the reference mode shapes shown in Figure \ref{fig:bridge} (c). Similar to the numerical experiments, some spurious modes appear in Figure \ref{fig:decomposition_laboratory}. However, they can be easily identified by examining their PSD magnitude and mode shape characteristics.

To comprehensively evaluate the trained model in modal identification, Table \ref{tab:identification_laboratory} provides the identified modal frequencies, damping ratios, and MAC values between the identified mode shapes and reference mode shapes obtained by finite element analysis. The results lead to the following observations: \textbf{1)} The modal identification results are satisfactory for Tests 1 to 8, which were used for model training, indicating that the model effectively learns the modal characteristics from these datasets. \textbf{2)} For tests where the excitation is a vehicle load, the modal identification performance declines, particularly for higher-order mode shapes and damping ratios. This is expected, as vehicle-induced excitation is nonstationary and non-Gaussian, making modal identification more challenging compared to the pull-and-release method. \textbf{3)} Regarding Tests 9 and 10, which are not included in model training, the trained model performs well on Test 9 but struggles with Test 10. This suggests that the model generalizes well to simpler cases like Test 9, where excitation follows the pull-and-release method, but its generalization ability is limited in more complex scenarios, such as Test 10, where the excitation source is a vehicle load.

\begin{table}[!htp] 
\centering
\caption{Modal identification results of laboratory experiments}
\label{tab:identification_laboratory}
\begin{tabular}{ccccccccccc}
\hline
         &            & \multicolumn{3}{c}{MAC}  & \multicolumn{3}{c}{Frequency (Hz)} & \multicolumn{3}{c}{Damping ratio (\%)} \\ \hline
Test No. & Excitation & mode 1 & mode 2 & mode 3 & mode 1   & mode 2   & mode 3  & mode 1      & mode 2      & mode 3     \\ \hline
1        & release    & 0.998  & 0.985  & 0.921  & 3.906    & 6.348    & 11.719  & 1.597       & 1.294       & 0.924      \\
2        & release    & 0.997  & 0.986  & 0.918  & 3.906    & 6.348    & 11.719  & 1.589       & 1.334       & 0.905      \\
3        & release    & 0.998  & 0.985  & 0.918  & 3.906    & 6.348    & 11.719  & 1.589       & 1.328       & 0.914      \\
4        & vehicle    & 0.985  & 0.982  & 0.827  & 4.004    & 6.543    & 11.719  & 2.141       & 1.508       & 1.832      \\
5        & vehicle    & 0.958  & 0.970  & 0.816  & 3.711    & 6.250    & 11.719  & 2.998       & 2.425       & 2.274      \\
6        & vehicle    & 0.979  & 0.983  & 0.906  & 4.004    & 6.543    & 11.816  & 1.290       & 1.864       & 1.585      \\
7        & vehicle    & 0.998  & 0.986  & 0.912  & 4.102    & 6.445    & 11.719  & 1.804       & 1.343       & 1.851      \\
8        & release    & 0.997  & 0.974  & 0.895  & 4.102    & 6.445    & 12.109  & 1.521       & 1.355       & 1.186      \\ \hline
9        & release    & 0.998  & 0.946  & 0.851  & 4.004    & 6.445    & 12.109  & 1.946       & 1.341       & 1.209      \\
10       & vehicle    & 0.390  & 0.697  & 0.327  & 4.004    & 6.445    & 12.695  & 3.161       & 0.356       & 2.042      \\ \hline
\end{tabular}
\end{table}

\subsection{Comparison against Classic Modal Identification Methods}
We further compare the modal identification results of our approach with those of Automated EFDD and SSI. This comparison is conducted solely on test 1-8, which are used for modal training. It should be again noted that, in this study, modal training inherently involves modal decomposition and identification, so this comparison is reasonable. For mode shape identification with Automated EFDD and SSI, these two methods are first used to determine mode shapes based on data from accelerometers A1 to A6. The identified mode shapes are then interpolated to estimate mode shapes at all nodes of the beam.

\begin{figure}[!htp] 
   \centering 
   \begin{tabular}{ccc}
   \includegraphics[width=5.2cm]{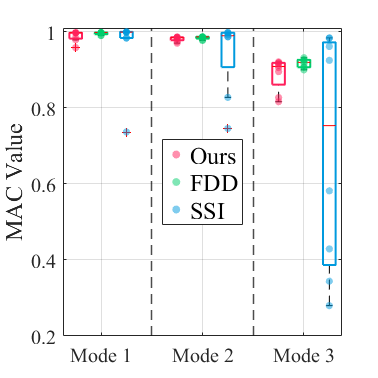} & \includegraphics[width=5.2cm]{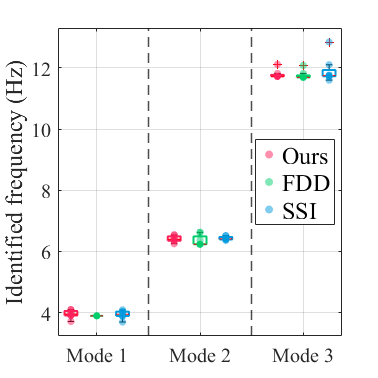} & \includegraphics[width=5.2cm]{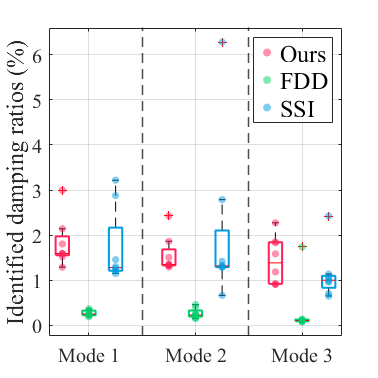} \\
   (a) & (b) & (c) \\
   \end{tabular}
    \caption{Modal identification results for tests 1-8, presented using box plots and scatter plots: (a) MAC values of the identified mode shapes, (b) Identified natural frequencies, and (c) Identified damping ratios.}
   \label{fig:comparison_laboratory}
\end{figure}

Figure \ref{fig:comparison_laboratory} visualizes the modal identification results for our approach, Automated EFDD, and SSI. Based on these results, the following observations can be made: \textbf{1)} For mode shape identification, both our approach and Automated EFDD achieve good performance, with EFDD showing slightly better accuracy. However, SSI produces some inaccurate results, which is expected, as SSI is known to perform poorly on short-duration signals, and signals in this study only last 10 seconds.\textbf{ 2)} For frequency identification, all three methods deliver comparable results, with no significant differences. \textbf{3)} For damping ratio identification, Our approach provides similar mean damping ratios to SSI, but with lower variance, indicating more stable results. In contrast, Automated EFDD tends to underestimate damping ratios, which is reasonable, as EFDD lacks the capability to accurately determine damping ratios. \textbf{4)} Overall, our approach achieves higher accuracy in comprehensive modal identification compared to EFDD and SSI.

Nonetheless, , our approach requires more time for modal decomposition and identification due to the computational demands of deep learning training. Specifically, the total time needed for modal identification across 10 tests is 2.322 seconds for EFDD, 27.183 seconds for SSI, and 304.005 seconds for our approach. However, since real-time modal identification is typically not a critical requirement, the extended computation time does not pose a major limitation. Furthermore, this additional cost is justified by the distinct advantage of our approach, which enables the separation of dynamic responses for individual structural vibration modes, a capability that neither EFDD nor SSI can achieve.

\section{Conclusions}
This study introduces an innovative approach that integrates Graph Neural Networks (GNNs), Transformers, a physics-informed loss function, and the Feature Propagation algorithm to conduct modal decomposition and identification across a structural population under realistic conditions. To evaluate the effectiveness of the proposed method, both numerical simulations and laboratory experiments are conducted. The key findings are summarized as follows:

1) \textbf{Practicality}: The physics-informed loss function, which leverages modal decomposition theory and the independence of structural vibration modes, enables the proposed deep learning model to be trained without requiring any labeled data. Additionally, the Feature Propagation algorithm allows for the identification of mode shapes at all structural nodes based on dynamic measurements from only a limited number of nodes. These two measures significantly enhance the practicality of the proposed approach.

2) \textbf{Effectiveness}: Once trained, the proposed deep learning model can effectively perform modal decomposition and identification for a population of structures used in model training, even in the presence of non-stationary conditions and limited dynamic measurements. Compared to conventional modal identification methods, such as Enhanced Frequency Domain Decomposition (EFDD) and Stochastic Subspace Identification (SSI), the proposed approach demonstrates comparable accuracy in identifying frequencies and damping ratios while achieving superior accuracy in identifying high-order mode shapes, attributed to its use of structural topology information. Moreover, unlike EFDD and SSI, the proposed approach is capable of providing dynamic responses of decomposed Single-Degree-of-Freedom (SDOF) vibration modes. In comparison to other deep learning models, including Multi-Layer Perceptrons (MLP) and Long Short-Term Memory (LSTM) networks, the GNN-Transformer combination yields significantly better performance, establishing it as a powerful tool for Population-Based Structural Health Monitoring.

3) \textbf{Limitations}: Although the proposed approach demonstrates strong generalization to new dynamic measurements of the same structure under simple excitations (e.g., free decay vibration), its generalization to new dynamic measurements under more complex excitations (e.g., vehicle loads) and to different structures remains limited. Consequently, model retraining is required for each new case, leading to substantially longer computation times compared to classical modal identification methods. While real-time modal decomposition and identification is typically unnecessary, future research should focus on enhancing the model’s generalization capabilities by incorporating additional physics-based constraints or refining the model architecture. This would enable accurate modal decomposition for new dynamic measurements solely through model inference, making the process significantly faster than traditional modal identification methods.



\printcredits

\section*{Declaration of Competing Interest}
The authors declare no competing interests exist.

\section*{Data Availability Statement}
Data and demonstrative Python codes that implement the proposed approach are openly available at our public GitHub repository: 
\href{https://github.com/JxdEngineer/ModalGNN_Time_Domain}{https://github.com/JxdEngineer/ModalGNN\_Time\_Domain}.

\section*{Acknowledgments}
The research was conducted at the Singapore-ETH Centre, which was established collaboratively between ETH Zurich and the National Research Foundation Singapore. The authors sincerely appreciate the support from the National Research Foundation, Prime Minister’s Office, Singapore under its Campus for Research Excellence and Technological Enterprise (CREATE) program.

\bibliographystyle{model1-num-names}

\bibliography{cas-refs}


\end{document}